\documentclass[final]{siamltex}
\usepackage{epsfig,amsmath,amssymb}

\newcommand{\E}{\mathbb{E}}

\title{Explosive behavior in a log-normal interest rate model}

\author{Dan Pirjol\thanks{J.~P.~Morgan, 277 Park Avenue, New York, NY 10172,
dpirjol@gmail.com}}

\begin{document}

\maketitle

\begin{abstract}
We consider an interest rate model with log-normally distributed rates 
in the terminal measure in discrete time. Such models are used in financial
practice as parametric versions of the Markov functional model, or as  
approximations to the log-normal Libor market model. We show that
the model has two distinct regimes, at low and high volatility,
with different qualitative behavior. The two regimes are separated by a sharp 
transition, which is similar to a phase transition in condensed matter physics.
We study the behavior of the model in the large volatility phase, and discuss the
implications of the phase transition for the pricing of interest rates derivatives.
In the large volatility phase, certain expectation values and convexity 
adjustments have an explosive behavior. 
For sufficiently low volatilities the caplet smile is log-normal to a very
good approximation, while in the large volatility phase the model
develops a non-trivial caplet skew.
The phenomenon discussed here imposes thus an upper limit on the 
volatilities for which the model behaves as intended.
\end{abstract}



\pagestyle{myheadings}
\thispagestyle{plain}

\section{Introduction}
An important class of interest rate models used in financial practice is 
the class of short rate models \cite{AP,1,Pelsser,BM}.
These models are Markovian, and the state of the 
model at time $t$ is completely defined by the short rate $r(t)$. 
As a result all dynamical variables such 
as zero coupon bonds and rates at some time $t$ depend only on $r(t)$. 
These models are simple and intuitive, but have the drawback that 
the connection between the model parameters and market data is not 
always transparent. For this reason
they are usually not flexible enough in their calibration to market data.

This problem can be solved by the introduction of market models. One particular
type of these models are the Markov functional models, where the dynamical
quantities of the model are functionals of a small number of Markov drivers
$x_i(t)$.
It has been shown in \cite{HKP,1,AP,BH} that by a judicious choice of the 
functional dependence of some dynamical variable of the model on the
stochastic drivers $x_i(t)$, it is possible to 
reproduce exactly Black's formula or a given market caplet or swaption 
implied volatility smile.

We consider in this paper a one-factor Markovian model in discrete time
with log-normally distributed rates in the terminal measure. Such a model 
is encountered in practice as a particular parametric realization of a 
Markov-functional model for simulating an interest rate market with log-normal 
caplet smile, see e.g. \cite{AP}. Models of this type
have been proposed in the literature as approximations to the 
Libor market model with log-normal caplet volatility \cite{BGM,LMMLN1,LMMLN2}.
For example, the Libor market model reduces to such a model in the frozen
drift approximation (up to the addition of appropriate convexity multipliers)
when considering only the terminal distribution of the
Libors at their setting time. See Sec.~\ref{sec:2} for a detailed discussion.

The model with log-normally distributed rates in the terminal measure
was studied in Ref.~\cite{PhaseTransition}, where it was shown that it 
can be solved exactly for the time-homogeneous case of uniform volatility. 
Exact results can be found for the dependence of all zero coupon bonds on 
the Markovian driver. 
The more general case of a short rate model where the short rate is the 
exponential of a Gaussian Markov process was considered in \cite{LatticeGas}.
The exact solution of the model was found to have discontinuous dependence
on volatility. The main results can be summarized as follows: 
the model has two distinct regimes, at low and large volatility, respectively, 
with different qualitative properties. These regimes
are separated by a sharp transition, occurring at a critical value of the 
volatility, which resembles a first order phase transition in condensed matter 
physics. In particular certain convexity adjustments appearing in the 
calibration of the model have non-analytic behavior as function of
the volatility, manifested as a discontinuous derivative at the critical
point.

In this paper we consider the implications of these phenomena for the pricing
of interest rates derivatives under such models, and point out that they 
limit their region of applicability.
We start by studying the distributional properties of the Libor probability 
distribution function in a measure where it is simply related to caplet prices
(the forward measure). It turns out that the shape of the Libor distribution
function changes suddenly at the critical volatility, and becomes very 
concentrated at small values of the Libor rates in addition to developing 
a long tail. This is reflected in Black caplet volatility, which undergoes a 
sudden change at the critical volatility, both in the ATM volatility and the 
shape of the caplet smile.
While for subcritical volatility the caplet
smile is flat to a good approximation, and is equal to the Libor volatility,
above the critical volatility the ATM caplet implied volatility increases
suddenly, and a nontrivial caplet smile appears.
In addition, the moments of the Libor probability distribution
function have non-analytic behavior in volatility and explosive behaviour.

Section~\ref{sec:2} introduces the model with log-normally distributed rates in the 
terminal measure which is considered in this paper, and provides some background on the 
practical relevance of such models. This section gives also a brief review of the main 
results of \cite{PhaseTransition}, describing the method used to find the exact 
solution of the model, and the main features of the volatility dependence following 
from this solution. In Section~\ref{sec:3} we study the
properties of the Libor probability distribution function in the forward measure
and its moments, and in Section~\ref{sec:4} we consider in some detail the pricing
of a Libor payment in arrears, and show the presence of
non-analyticity in prices of actual interest rate derivatives in this model. 
Section~\ref{sec:5} gives a summary of the main results, and a brief discussion of their
implications. 

\section{The model}
\label{sec:2}

We consider an interest rate model defined on a finite set of dates $t_i$
\begin{eqnarray}
0 = t_0 < t_1 < \cdots < t_n\,.
\end{eqnarray}
Typically the dates $t_i$ are equally spaced, e.g. by 3 or 6 months apart, but we 
will keep them completely general, and denote the difference between
consecutive dates as $\tau_i \equiv t_{i+1} - t_i$.

The fundamental dynamical quantities of the model are the zero coupon bonds 
$P_{i,j}\equiv P_{t_i,t_j}$. They are driven by a one-dimensional Markov 
process $x(t)$, which will be assumed to be a simple Brownian motion under
the terminal measure. The numeraire is $P_{t,t_n}$, the zero coupon bond
maturing at the last simulation time $t_n$. Define the forward Libor
rate $L_i(t)$ for the $(t_i,t_{i+1})$ period as
\begin{eqnarray}
L_{i}(t) = \frac{1}{\tau_i} \Big(
\frac{P_{t,t_i}}{P_{t,t_{i+1}}} - 1\Big) \,.
\end{eqnarray}

The model is defined by the functional dependence on $x(t)$ of the Libor 
rates at their setting time $L_{i} \equiv  L_i(t_i)$.
Specifically, in the model considered here the Libors $L_{i}$ are assumed 
to be log-normally distributed in the terminal measure, 
\begin{eqnarray}\label{Lipdf}
L_i = \tilde L_i \exp\Big( \psi_i x_i - \frac12 \psi_i^2 t_i \Big) \,.
\end{eqnarray}
For simplicity we will denote the value of the Markov driver at time $t_i$ 
as $x_i \equiv x(t_i)$. Its mean and variance are $\mathbb{E}[x_i]=0,
\mathbb{E}[x_i^2]=t_i$. Here $\psi_i$ is the volatility of the Libor $L_i$,
and $\tilde L_i$ are parameters which will have be determined such that the
initial yield curve $P_{0,i}$ is correctly reproduced. 

The original motivation for this work was the study of a Markov-functional
model with log-normal functional specification of the Libors at their setting
time on the Markovian driver, as given by (\ref{Lipdf}).
Such a specification is somewhat different from the original philosophy 
behind the Markov-functional models \cite{HKP,1,AP,BH}, which aims to
reproduce exactly Black's formula or model an appropriate 
market skew for caplets or for swaptions with a judicious
choice for the functional dependence $L_i(x_i)$ (or equivalently the
functional dependence of the numeraire $P_{t,t_n}(x)$).
See \cite{Johnson} for a detailed discussion of a specific implementation.

The Markov-functional model is used in practice in parametric and 
non-parametric versions, according to the implementation of the
functional dependence $L_i(x_i)$. The log-normal parametric form considered
here is one of the simplest possible, and was considered first in 
the original work on Markov functional models \cite{HKP}. For another
discussion of this model see Sec.~11.A.2 in \cite{AP}.
Another parametric version is the so-called
semi-parametric representation of Ref.~\cite{Pelsser}
\begin{eqnarray}
P^{-1}_{i,n}(x) = 1 + a_i e^{b_i x} + d_i e^{-\frac12 c_i(x - m_i)^2}
\end{eqnarray}
where $a_i,b_i,c_i,d_i,m_i$ are numerical parameters to be fitted to 
the numerical solution. 
Of course, with a parametric representation the resulting model will not 
reproduce exactly the caplet or swaption implied volatility for all strikes,
and the best one can aim for is to match the implied volatility at one
particular strike (for example the ATM point), and possibly also for a 
region of neighboring strikes.

Models with log-normal rates in the terminal measure appear also when
considering approximations to the log-normal Libor market model (LMM) \cite{BGM}. 
For a general comparison of the LMM with separable Libor volatility and
Markov-functional models in the one-dimensional case see \cite{BeKe}.
Consider a model with log-normal caplet volatilities $\psi_i$ and given
initial yield curve $P_{0,i}$ implying forward Libors $L_i^{\rm fwd} = 1/\tau_i(P_{0,i}/P_{0,i+1}-1)$. 
A one-factor LMM realization of this model is specified by the diffusion
for the forward Libors $L_i(t)$  \cite{BGM}
\begin{eqnarray}
\frac{dL_i(t)}{L_i(t)} = \psi_i dx(t) - \sum_{j=i+1}^{n-1}
\frac{\tau_j \psi_j^2 L_j(t)}{1+L_j(t)\tau_j} dt
\end{eqnarray}
where $x(t)$ is a Brownian motion in the terminal measure. 
This model is usually simulated using Monte-Carlo methods \cite{AP}, 
but several
analytical approximations have been proposed as well. The simplest
approximation is the frozen drift approximation, where the
forward Libors $L_j(t)$ in the drift term are replaced with their forward
values $L_j(0) = L_j^{\rm fwd}$. Then the evolution equation can be
solved in closed form, and the Libors at their setting time are
log-normally distributed in the terminal measure
\begin{eqnarray}
L_i^{\rm FD}(t_i) &=& L_i^{\rm fwd} \exp\Big( \psi_i x_i - \frac12 \psi^2 t_i
+ \mu_i t_i\Big) \\
\mu_i &=& - \sum_{j=i+1}^{n-1}
\frac{\tau_j \psi_j^2 L_j^{\rm fwd}}{1+L_j^{\rm fwd}\tau_j}
\end{eqnarray}
This has the form of (\ref{Lipdf}) with $\tilde L_i = L_i^{\rm fwd} 
e^{\mu_i t_i}$. This expression for $\tilde L_i$ agrees with the
small-volatility limit of the convexity-adjusted Libors $\tilde L_i$
derived in Eq.~(27) of \cite{PhaseTransition}. 
For larger volatilities a convexity adjuster $\kappa_i$ must be added 
to the frozen drift approximation $\tilde L_i = \kappa_i L_i^{\rm fwd} 
e^{\mu_i t_i}$ such that the yield curve $P_{0,i}$ is correctly reproduced.
The frozen drift approximation is expected to be valid only for very
small volatility. 
Improved approximations to the Libor market model which
are valid over a larger range of volatilities, and which have
log-normally distributed Libors in the terminal measure have been
proposed in \cite{LMMLN1,LMMLN2}.

This paper presents a study of the model defined by (\ref{Lipdf}) 
with uniform volatility parameter\footnote{The restriction to the case of 
uniform $\psi_i$ is motivated by arguments of simplicity of the resulting 
analytical formulas. The general case of arbitrary $\psi_i$ can be also
solved in closed form \cite{LatticeGas}, but the resulting expressions
are unwieldy for numerical evaluation for $n > 12$, with $n$ the number of 
simulation times.} 
$\psi_i = \psi$, and investigates its properties as functions of the volatility 
parameter $\psi$.
For sufficiently low volatility the model is found to generate a log-normal
caplet smile, with caplet volatility equal to $\psi$ to a very 
good approximation. In retrospect, this provides a justification for the
choice of the functional form (\ref{Lipdf}) for describing a model
with log-normal caplet volatility. In other words, for sufficiently
small caplet volatility, a model with log-normal caplet smile has also
log-normally distributed rates in the terminal measure to a very 
good approximation.

As the volatility parameter $\psi$ is increased, the ATM caplet volatility
starts to diverge from the model volatility parameter $\psi$, and a 
non-trivial caplet smile appears. This change is not gradual, but rather
occurs at a sharply defined value of the volatility parameter $\psi_{\rm cr}$,
which we call the critical volatility. We will show that at the critical
volatility the ATM caplet volatility
has a sudden increase, and the shape of the Libor probability distribution in 
its natural measure changes suddenly from a typical humped shape to a function 
which is very concentrated near the origin.

Above the critical volatility, the dynamics predicted from the model are 
different from those intended (log-normal caplet smile), such that this phenomenon 
introduces a limitation of the model, or more precisely of the choice (\ref{Lipdf})
for the functional dependence of the Libor distribution. In the context of the
Libor market model, our results give a measure of the limit of validity of a
log-normal approximation. In view of the practical use of this parameterization, 
it is important to understand its region of applicability.

\subsection{Analytical solution}

We summarize here the derivation and main features of the analytical solution 
of the model \cite{PhaseTransition}.
The zero coupon bonds $P_{i,j}(x)$ can be expressed as functions 
of the one-dimensional Markov process $x(t)$.
We will denote the numeraire-rebased zero coupon bond prices as
$\hat P_{i,j} = P_{i,j}/P_{i,n}$.
They are martingales in the terminal ($t_n$-forward) measure, and thus satisfy
the condition, see e.g. \cite{BR}
\begin{eqnarray}\label{martingale}
\hat P_{i,j} = \E_n\Big[\frac{1}{P_{j,n}} \Big| {\cal F}_i\Big]
\end{eqnarray}
Imposing the martingale condition (\ref{martingale}) for 
all possible $i,j$ pairs determines uniquely the convexity-adjusted 
Libors $\tilde L_i$, provided that the volatility parameters $\psi_i$
are known. The latter are  determined for example by calibration to the ATM 
caplet volatilities, observed in the market.

This model can be solved analytically \cite{PhaseTransition,LatticeGas}.
For the case of uniform volatility $\psi_i = \psi$ the solution can be expressed 
as an analytical expression for the one-step 
zero coupon bonds $\hat P_{i,i+1}$ 
\begin{eqnarray}
\hat P_{i,i+1}(x_i) = \sum_{j=0}^{n-i-1} c_j^{(i)} e^{j\psi x_i - \frac12 (j\psi)^2 t_i}
\end{eqnarray}
with $c_j^{(i)}$ a set of constant coefficients to be determined.
The convexity-adjusted Libors are given by
\begin{eqnarray}\label{LtildeNi}
&& \tilde L_{i} = \frac{\hat P_{0,i} - \hat P_{0,i+1}}{N_i\tau _i}\,,\\
&&  N_i \equiv \E[\hat P_{i,i+1} e^{\psi x_i -\frac12\psi^2 t_i}] 
= \sum_{j=0}^{n-i-1} c_j^{(i)} e^{j\psi^2 t_i} \nonumber\,.
\end{eqnarray}

The coefficients $c_j^{(i)}$ satisfy the recursion relation
\begin{eqnarray}\label{recursionci}
c_j^{(i)} = c_j^{(i+1)} + \tilde L_{i+1} \tau_{i+1}c_{j-1}^{(i+1)} e^{(j-1) \psi^2 t_{i+1}}
\end{eqnarray}
which must be solved simultaneously with Eq.~(\ref{LtildeNi}) for $\tilde L_i$. The initial
condition is $c_0^{(n-1)}=1, \tilde L_{n-1}\tau_{n-1} = \hat P_{0,n-1}-1$. The recursion relation (\ref{recursionci})
can be solved backwards in time, for all $i \leq n-1$, finding all coefficients
recursively. An explicit solution of this recursion relation was found in 
\cite{LatticeGas}, and $c_j^{(i)}$ has a physical interpretation as the canonical 
partition function of a one-dimensional attractive Coulomb lattice gas 
with $j$ particles and $n-i-1$ sites. We will not make use of this solution
here, and prefer to evaluate the recursion relation (\ref{recursionci}) explicitly.
The coefficients $c_j^{(i)}$ and the convexity-adjusted Libors $\tilde L_i$ determine the solution
of the model. The zero coupon bonds $P_{i,j}$ can be found
explicitly as functions of $x=x_i$, as shown in Eq.~(\ref{recursionci}).

The recursion relation (\ref{recursionci}) can be expressed more compactly
by introducing the generating function at the time horizon $t_i$
\begin{eqnarray}
f^{(i)}(x) \equiv \sum_{j=0}^{n-i-1} c_j^{(i)} x^j
\end{eqnarray}
The generating function $f^{(i)}(x)$ takes known values at $x=0,1$
\begin{eqnarray}\label{fi01}
f^{(i)}(0) = 1\,,\qquad
f^{(i)}(1) = \hat P_{0,i+1}
\end{eqnarray}
where the second constraint follows from a sum rule for the 
coefficients $c_j^{(i)}$ \cite{PhaseTransition}.
The generating function satisfies the recursion relation 
\begin{eqnarray}\label{firecursion}
f^{(i)}(x) = f^{(i+1)}(x) + \tilde L_{i+1} \tau x f^{(i+1)}(x e^{\psi^2 t_{i+1}})
\end{eqnarray}
with initial condition $f^{(n-1)}(x) = 1$.
The expectation value $N_i$ appearing in the expression for the convexity-adjusted Libor 
$\tilde L_{i}$ is
\begin{eqnarray}\label{Nif}
N_i = f^{(i)} (e^{\psi^2 t_i} )
\end{eqnarray}

The generating function $f^{(i)}(x)$ and thus the coefficients $c_j^{(i)}$
can be found in closed form in the two limiting
cases of very small and very large volatility $\psi$ \cite{PhaseTransition}. 
The zero volatility limit of the generating function is 
\begin{eqnarray}\label{fi0}
f^{(i)}_0(x) = \Pi_{j=i+1}^{n-1} (1 + L_{j}^{\rm fwd} \tau x)\,.
\end{eqnarray}

In the asymptotically large volatility limit $\psi \to \infty$, the
recursion relation (\ref{firecursion}) can be solved again exactly with the result
\begin{eqnarray}\label{fiasym}
f^{(i)}_\infty(x) = 1 + (\hat P_{0,n-1} - 1)x 
+ \cdots + (\hat P_{0,i+1} - \hat P_{0,i+2})x^{n-i-1} \,.
\end{eqnarray}
The most distinctive feature of the model in the large volatility limit is an
explosive increase of the expectation values $N_i = f^{(i)}(e^{\psi^2 t_i})$ 
with the volatility. This causes the convexity-adjusted Libors
$\tilde L_i$ to become very small. Their asymptotic expression in the large volatility
phase is \cite{PhaseTransition}
\begin{eqnarray}\label{Ltildei}
\tilde L_i = \frac{\hat P_{0,i} - \hat P_{0,i+1}}
{(\hat P_{0,i+1} - \hat P_{0,i+2})\tau_i} 
e^{-(n-i-1)\psi^2 t_i} (1 + O(e^{-\psi^2 t_i})) \,.
\end{eqnarray}
In practice $\tilde L_i$ can become very small, below machine precision, 
which can make an exact numerical simulation of the model very difficult 
in  the large volatility regime.

\subsection{The Libor phase transition}

The analytical solution of the model presented above can be used to study exactly its
behavior as a function of the volatility parameter $\psi$. It turns out that this is not 
smooth for all quantities of the model. Certain expectation values, such as $N_i$ given in 
(\ref{LtildeNi}), have a $\psi$ dependence which has singular behavior at a special value 
of volatility which will be called the critical volatility $\psi_{\rm cr}$. This is
manifested as a sudden change in the derivative $dN_i/d\psi$ at the critical point, which 
becomes more sharp as the time step $\tau$ decreases, such that it approaches a nonanalyticity
point in the continuous time limit \cite{PhaseTransition}. At the critical point
the expectation value $N_i$ has an explosive increase, which is much faster than in the
low volatility phase.

The underlying reason for this phenomenon is a singularity in the 
generating function $f^{(i)}(x)$ at a certain value $x_*$. This value is related to the
position of the zeros of $f^{(i)}(x)$ in the complex plane. The generating function is
a polynomial in $x$ of degree $n-i-1$ with positive coefficients, and thus does not
have any zeros on the positive real axis. It has $n-i-1$ zeros, which are arranged
in complex conjugate pairs symmetric with respect to the real axis, along a curve
surrounding the origin. The singularity point $x_*$ is the point on the positive real axis
where the complex zeros pinch the real axis. At this point the derivative of the generating
function has a discontinuity which is proportional to the angular density of the zeros around
the positive real axis. This density is of the order of $n-i-1$, the number of 
simulation time steps to the maturity.

This phenomenon is similar to a first order phase transition in condensed matter physics, 
where the thermodynamical potentials have a discontinuity in the first derivative at the
critical point \cite{EHS}. The analogy becomes even closer in the Lee-Yang formalism of the 
phase transitions \cite{LY}, where the critical point is associated with the complex zeros
of the grand canonical partition function.

In the context of the Markov functional model with log-normally distributed rates, the
singularity in $N_i$ occurs at the point $\psi_{\rm cr}$ given by
\begin{eqnarray}\label{volcr}
e^{\psi^2_{\rm cr} t_i} = x_* \,.
\end{eqnarray}
This equation determines the critical volatility $\psi_{\rm cr}$.
A similar phenomenon occurs for any expectation value of the form similar to $N_i$
\begin{eqnarray}
\mathbb{E}_n[\hat P_{i,i+1} e^{\phi x - \frac12 \phi^2 t_i}] = f^{(i)} (e^{\psi\phi t_i})
\end{eqnarray}
which can be expressed in terms of the generating function $f^{(i)}(x)$ as shown.
The critical volatility corresponding to this expectation value is found in analogy to
Eq.~(\ref{volcr}) and is given by $\exp(\psi\phi t_i) = x_*$.

The precise value of the critical volatility depends on the time step $i$, 
and on the entire shape of the yield curve $P_{0,i}$. Consider
for illustration the case of a constant short rate $r_0$, which corresponds
to the discount bonds $P_{0,i}=\exp(-r_0 \tau i)$. A simple estimate of the
critical volatility can be obtained from the zeros of the asymptotic generating
function $f_\infty^{(i)}(x)$, corresponding to very 
large volatility \cite{PhaseTransition}.
Using an approximation for the position of these zeros one finds
\begin{eqnarray}\label{r0cr}
e^{r_0 \tau + \psi_{\rm cr}^2 t_i} = \Big(
\frac{1}{1-e^{-r_0\tau}} \Big)^{1/(n-i-1)}
\end{eqnarray}
which can be approximated, to a good precision, as
\begin{eqnarray}\label{psicr}
\psi^2_{\rm cr} = \frac{1}{i(n-i-1)\tau}\log\Big(\frac{1}{r_0\tau}\Big)\,.
\end{eqnarray}

The relation (\ref{psicr}) reproduces the main features of the critical volatility 
observed in numerical simulations:
\begin{itemize}

\item The critical volatility decreases as the size of the time step $\tau$ is
reduced, approaching zero in the continuous time limit. 

\item The critical volatility increases as the short rate $r_0$ is reduced, 
approaching a very large volatility as the rate $r_0$ becomes very small.

\end{itemize}

\begin{figure}
\begin{center}
\includegraphics[height=40mm]{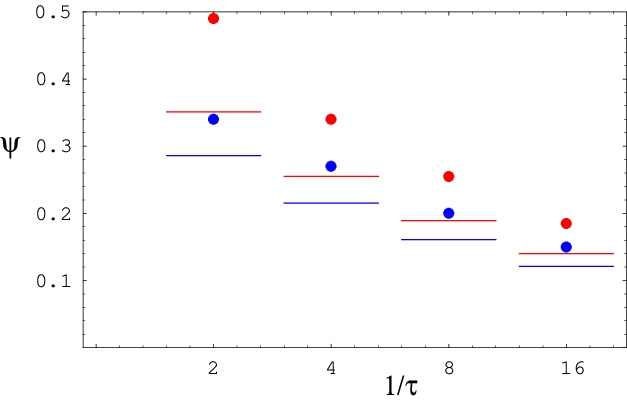}
\includegraphics[height=40mm]{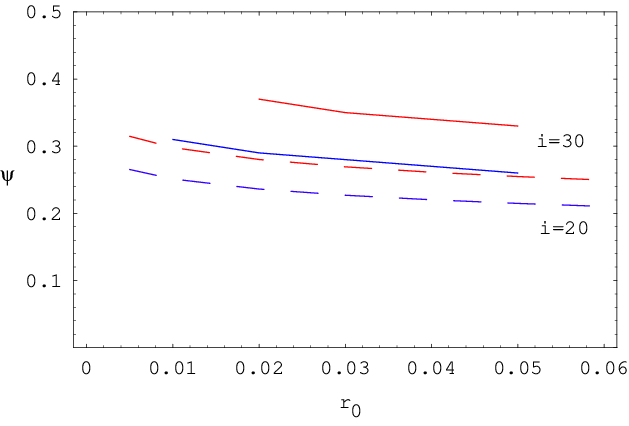}
\end{center}
\caption{Left: the critical volatility $\psi_{\rm cr}$ of the 
log-normal interest rate model with constant forward short rate $r_0=5\%$
for several discretizations with $1/\tau = 2,4,8,16$ time steps per year.
The dots show the exact critical volatility, and the lines the results of
the approximation (\ref{psicr}). Blue: Libor rate set at $t_i=5$, red:
Libor rate set at $t_i=7.5$.
Right: the dependence of the critical volatility $\psi_{\rm cr}(r_0)$
on the short rate  in a simulation with quarterly time steps $\tau=0.25$.
The red curves correspond to the Libor rate set at $t_i=7.5$ and the 
blue curves to the Libor rate set at $t_i=5$. The solid lines are
exact phase boundaries, while the dashed lines correspond to the
approximative result Eq.~(\ref{psicr}).
Total simulation time is $t_n=10$ years.}
\label{fig:phasediagram}
\end{figure}

This behavior is illustrated in Fig.~\ref{fig:phasediagram}.
These plots show the critical volatility $\psi_{\rm cr}$ of the model with 
flat forward short rate $r_0$ as function of $\tau$ at fixed $r_0$ 
(left panel),
and as function of $r_0$ at fixed time step $\tau$ (right panel). 
In these plots we show both the exact critical volatility (dots/solid lines), 
which can be found as the value of $\psi$ at which $\partial_\psi^2 \log N_i$ 
is maximal, and the result of the simple approximation (\ref{psicr}) 
(lines/dashed lines). These plots show that the approximation (\ref{psicr}) 
underestimates the actual value of the critical volatility by about 10\%.

\section{The Libor probability distribution function}
\label{sec:3}

By the model definition (\ref{Lipdf}), the Libor rates $L_i$ are log-normally 
distributed in the $t_n$-forward measure (the terminal measure). A more natural 
measure for pricing instruments depending on 
the Libor rate $L_i$ is the $t_{i+1}$-forward measure (or simply the forward
measure), with numeraire the
zero-coupon bond $P_{t,i+1}$, maturing at time $t_{i+1}$. We will consider 
the two measures
\begin{eqnarray}
&& \mathbb{P}_n \, : \quad\,\,\,\,\, \mbox{numeraire } P_{t,n} \\
&& \mathbb{P}_{i+1} \, : \quad \mbox{numeraire } P_{t,i+1} 
\end{eqnarray}

As a concrete example, consider a caplet ${\bf C}_i(K)$ on the Libor rate 
$L_{i} = \tau_i^{-1} (P_{i,i+1}^{-1} - 1)$, set at time $t_i$ and 
paid at time $t_{i+1}$, with strike $K$. The payoff of this instrument 
is $(L_i - K)_+$, and its price is given as an expectation value in the 
$t_n$-forward measure $\mathbb{P}_n$ 
\begin{eqnarray}\label{Capletn}
C_i(K) = P_{0,n} \E_n[(L_i - K)_+ \hat P_{i,i+1}] \,.
\end{eqnarray}

Expressed in the forward $\mathbb{P}_{i+1}$ measure, the expression for the
caplet price ${\bf C}_i(K)$ simplifies and is given by
\begin{eqnarray}\label{Capleti}
C_i(K) = P_{0,i+1} \E_{i+1}[(L_i - K)_+]
\end{eqnarray}
The expectation value in $\mathbb{P}_{i+1}$ measure can be expressed
as an integral of the payoff convoluted with the probability distribution
function of the Libor $L_i$
in this measure. We will denote this distribution $\Phi_i(L_i)$, 
and we have
\begin{eqnarray}
\E_{i+1}[(L_i - K)_+] = \int_0^\infty dL_i \Phi_i(L_i) (L_i - K)_+
\end{eqnarray}

In the following we will study in some detail the distribution function
$\Phi_i(x)$, and its properties. 
The pdf of the Libor $L_i$ in the $\mathbb{P}_{i+1}$ measure can be obtained
by comparing Eqs.~(\ref{Capletn}) and (\ref{Capleti}). It is given by
\begin{eqnarray}\label{PhiL}
\Phi_i(L) = \frac{1}{\hat P_{0,i+1}}
\frac{e^{-x_0^2/(2t_i)}}{\sqrt{2\pi t_i}} \frac{1}{\psi L} \hat P_{i,i+1}(x_0)
\end{eqnarray}
with $x_0 = x_0(L)$ determined as
\begin{eqnarray}\label{x0L}
x_0 = \frac{1}{\psi}\log \frac{L}{\tilde L_i} + \frac12 \psi t_i\,.
\end{eqnarray}

\begin{figure}
\begin{center}
\includegraphics[height=50mm]{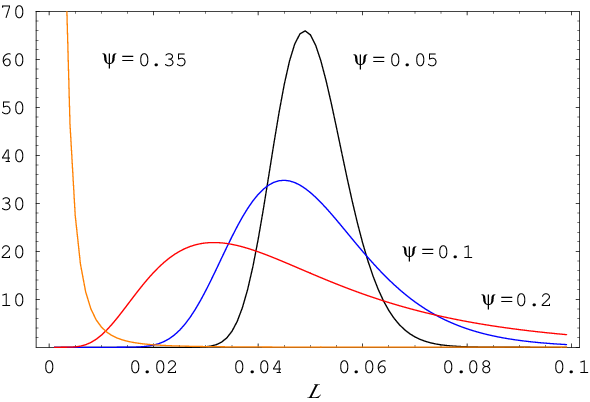}
\end{center}
\caption{The probability distribution function $\Phi_i(L)$ for the Libor $L_i$
in the measure $\mathbb{P}_{i+1}$ for several values of the volatility
$\psi$. The plots correspond to a constant forward short rate $r_0=5\%$,
which gives $L_i^{\rm fwd}=5.0314\%$. The remaining parameters are $i=30$,
$n=40$, $\tau = 0.25$.}
\label{fig:pdfpsi}
\end{figure}

We would like to study how the Libor probability distribution
function $\Phi_i(L)$ changes as the volatility $\psi$ is increased from zero
to large values. At zero volatility $\psi = 0$, this distribution is
a delta function concentrated at the forward value
\begin{eqnarray}
\Phi_i(L,\psi =0 ) = \delta(L - L_i^{\rm fwd})
\end{eqnarray}
As the volatility increases, the distribution widens out. We show in Fig.~\ref{fig:pdfpsi}
the shape of the distribution $\Phi_i(L)$ for several values of the volatility $\psi$. 
For moderate values of $\psi$, below the critical volatility $\psi_{\rm cr}$, the 
distribution has a typical humped shape, peaked around the forward value $L_i^{\rm fwd}$.

Above the critical volatility $\psi > \psi_{\rm cr}$ the probability distribution function 
$\Phi_i(L)$ undergoes a dramatic change: its support appears to collapse 
very rapidly to very small values of $L$, see Fig.~\ref{fig:pdfpsi}.
The ``collapse'' of the support of the Libor distribution $\Phi_i(L)$ to very small values 
close to zero is another surprising phenomenon in the high-volatility phase of 
this model.

Naively, one may ascribe this phenomenon to the fact that the convexity-adjusted
Libors $\tilde L_i$ in the defining equation of the model (\ref{Lipdf}) become very small 
in the large volatility phase. Upon further reflection the situation is 
slightly more complicated, for two reasons.
First, the log-normal distribution (\ref{Lipdf}) is in the terminal measure $\mathbb{P}_n$,
while we are interested here in the probability distribution function in $\mathbb{P}_{i+1}$ 
measure. 
Second, the martingale condition for $L_i(t)$ in $\mathbb{P}_{i+1}$ measure requires that the
average of $L_i(t_i)$ should be equal to its forward value $\E_{i+1}[L_i] = L_i^{\rm fwd}$, which would 
not be possible if the distribution were concentrated near $L_i=0$. The only way for the 
martingale condition to be satisfied is that the distribution has a long fat tail, which
contributes significantly to the average of $L_i$.

In the following we would like to explore this phenomenon in more detail. The
analysis presented next will confirm the heuristic arguments mentioned above.
We start by showing that the probability distribution function $\Phi_i(L)$ can be represented as
a sum of log-normal distributions.
Define the log-normal distribution with average parameter $\mu$ and dispersion $\sigma$
\begin{eqnarray}
\phi(x;\mu,\sigma) = \frac{1}{x\sqrt{2\pi} \sigma} e^{-\frac{1}{2\sigma^2}(\log x - \mu)^2}
\end{eqnarray}
The $j$-th moment of the $x$ variable under the distribution $\phi(x;\mu,\sigma)$ is
\begin{eqnarray}
\E[x^j] = e^{j\mu + \frac12 j^2 \sigma^2}
\end{eqnarray}

The probability distribution function of the $L_i$ Libor in the $\mathbb{P}_{i+1}$ measure (\ref{PhiL})
can be represented as a sum of log-normal distributions with different averages but the same variance
\begin{eqnarray}\label{mix}
\Phi_i(L) = \frac{1}{\hat P_{0,i+1}}\sum_{j=0}^{n-i-1} c_j^{(i)} \phi(L; \mu_j^{(i)}, \sigma_i = \psi \sqrt{t_i} )
\end{eqnarray}
where
\begin{eqnarray}
\mu_j^{(i)} = \log \Big( \tilde L_i e^{(j-\frac12)\psi^2 t_i}\Big) \,.
\end{eqnarray}
The average value of $L$ under the log-normal distribution $\phi(L; \mu_j^{(i)}, \sigma_i = \psi \sqrt{t_i} )$ is 
\begin{eqnarray}
\E[ L|\phi(L;\mu_j^{(i)},\sigma_i )] = \tilde L_i  e^{j\psi^2 t_i}
\end{eqnarray}
so each of these log-normal distributions are peaked at successively higher values of $L$. 
Specifically, the pdf of the Libor (\ref{mix}) consists of a sum of log-normal distributions with averages $\tilde L_i, \tilde L_i e^{\psi^2 t_i},
 \cdots, \tilde L_i e^{(n-i-1)\psi^2 t_i}$, and weights $c_k^{(i)}/\hat P_{0,i+1}$
with $k=0,1,\cdots, n-i-1$.
We recall that the
weights add up to 1 due to the exact sum rule $\sum_{j=0}^{n-i-1} c_j^{(i)} = \hat P_{0,i+1}$.

The weights $c_j^{(i)}/\hat P_{0,i+1}$ of the terms with $j > 1$ decrease sufficiently fast with $j$, such that the total average of $L$ is equal to the
forward Libor rate, as required by the martingale condition for $L_i$ in the $\mathbb{P}_{i+1}$ measure
\begin{eqnarray}\label{nonarb}
\mathbb{E}_{i+1}[L] &=& \frac{1}{\hat P_{0,i+1}} \sum_{j=0}^{n-i-1} c_j^{(i)} 
\mathbb{E}[ L|\phi(L;\mu_j^{(i)},\sigma_i )]\\
&=& \frac{1}{\hat P_{0,i+1}} \sum_{j=0}^{n-i-1} c_j^{(i)} 
\tilde L_i e^{j\psi^2 t_i} = L_i^{\rm fwd}\,.\nonumber
\end{eqnarray}

The representation (\ref{mix}) of the distribution function can be used 
to obtain a qualitative understanding of the behavior of this 
function in the large volatility limit $\psi^2 t_i \gg 1$. In this limit the 
asymptotic behavior of the convexity adjusted Libors $\tilde L_i$ is given by
(\ref{Ltildei}).
In the large volatility regime, the convexity adjusted Libors decrease very rapidly with 
the volatility $\psi$. 
This means that most of the log-normal components of the distribution function $\Phi_i(L)$ have 
vanishingly small averages, except for the last one with
the largest index $j=n-i-1$
\begin{eqnarray}
\mathbb{E}[ L|\phi(L;\mu_j^{(i)},\sigma_i )] &=& L_i^{\rm max} e^{-(n-i-1-j)\psi^2 t_i}\,,
\end{eqnarray}
where
\begin{eqnarray}
L_i^{\rm max} = \frac{\hat P_{0,i} - \hat P_{0,i+1}}
{(\hat P_{0,i+1} - \hat P_{0,i+2})\tau_i} =
L_i^{\rm fwd} \frac{1+L_{i+1}^{\rm fwd}\tau_{i+1}}{L_{i+1}^{\rm fwd}\tau_{i+1}}\,.
\end{eqnarray}
For typical values of model parameters, such as $L^{\rm fwd} =  5\%, \tau = 0.25$, one has
$L^{\rm max} \sim 500\%$ which is a very large value compared to typical rates.

In the large volatility regime the coefficients $c_j^{(i)}$ are all comparable, such that the 
shape of the probability distribution function $\Phi_i(L)$ is
expected to be very concentrated near $L=0$, corresponding to the terms with $j=0,1,\cdots, n-i-2$, 
and to have a fat tail extending to very large values of $L \sim L^{\rm max}$, 
corresponding to the term with $j=n-i-1$.
This is confirmed by direct calculation of the distribution function 
in the large volatility limit, as observed in Fig.~\ref{fig:pdfpsi}.

The behavior of the Libor probability distribution function in the 
$\mathbb{P}_{i+1}$ measure in the 
large volatility limit is related to a numerical issue discussed in \cite{PhaseTransition},
which is responsible for the unobservability of the phase transition in usual simulation
methods such as Monte Carlo or finite difference methods. This numerical issue appears
in the calculation of the expectation value (\ref{LtildeNi}) as an integral
\begin{eqnarray}\label{NiIntegrand}
N_i = \E[\hat P_{i,i+1} e^{\psi x - \frac12 \psi^2 t_i}] 
= \int_{-\infty}^\infty \frac{dx}{\sqrt{2\pi t_i}} e^{-\frac{x^2}{2t_i}}
\hat P_{i,i+1}(x) e^{\psi x - \frac12 \psi^2 t_i} \,. 
\end{eqnarray}
At volatilities above the critical value $\psi > \psi_{\rm cr}$  
the integrand in this expression develops a secondary peak at a relatively large 
value of $|x|\sim 10\sqrt{t_i}$, in addition to the peak around $x\sim 0$,
see Fig.~4 in \cite{PhaseTransition}. 
The secondary peak gives the dominant contribution to the integral in the 
large volatility limit. However, the region of large $x$ where it appears is 
either very poorly sampled, or completely ignored in usual simulation methods, which 
thus will fail to take it into account.

The dominance of the integral by the secondary peak in the super-critical regime
can be understood by changing variables in the integral (\ref{NiIntegrand}) 
from $x$ to $L_i$, the Libor rate.
We observe that the integrand of (\ref{NiIntegrand}) is simply related to the
Libor probability distribution function $\Phi_i(L)$ in the $\mathbb{P}_{i+1}$ measure,
when expressed in terms of $x = x(L)$ given in Eq.~(\ref{x0L}). 
The integral in (\ref{NiIntegrand}) becomes, after  changing the integration 
variable from $x$ to $L$
\begin{eqnarray}\label{LiNA}
N_i = \hat P_{0,i+1} \int_0^\infty dL \Phi_i(L) \frac{L}{\tilde L_i}\,.
\end{eqnarray}
The secondary peak in the integrand of (\ref{NiIntegrand}) becomes the fat tail of 
$\Phi_i(L)$, while the peak near $x\sim 0$ corresponds to the region of $L\sim \tilde L_i$. 
As discussed above, the fat tail of $\Phi_i(L)$ is essential in order for the
integral above to reproduce correctly its non-arbitrage value; the counterpart 
of this statement in the $x-$integral (\ref{NiIntegrand}) is that the secondary 
peak is also required by the consistency of the model, and can not be neglected.

\subsection{The moments of the Libor pdf}
\label{sec:3.1}

In this section we consider the moments of the Libor probability distribution 
function $\Phi_i(L)$ in the forward $\mathbb{P}_{i+1}$ measure. We will show that its 
moments, and thus its characteristic function
\begin{eqnarray}
\tilde \Phi_i(u) &=& \int_{-\infty}^\infty dL e^{iu L} \Phi_i(L)  
=\sum_{j=0}^\infty  \frac{(iu)^j}{j!} \mathbb{E}_{i+1}[L_i^j]
\end{eqnarray}
can be expressed in terms of the generating function $f^{(i)}(x)$. 

The moments of $\Phi_i(L)$ can be computed using Eq.~(\ref{mix}). 
The integral can be performed straightforwardly by changing the integration
variable from $L$ to $x$. The result expresses the $j-$th moment of the
Libor distribution in the forward measure in terms of the generating
function $f^{(i)}(x)$ as
\begin{eqnarray}\label{momL}
M_j = \mathbb{E}_{i+1}[L_i^j] &=& \int_0^\infty dL_i (L_i)^j \Phi_i(L_i)
 = \frac{1}{\hat P_{0,i+1}}
{\tilde L_i}^j e^{\frac12 j(j-1)\psi^2 t_i} f^{(i)}(e^{j\psi^2 t_i})\,.
\end{eqnarray}

We consider a few particular cases of the relation Eq.~(\ref{momL}). 
The first two moments $j=0,1$ do not contain dynamical information, and are
constrained by general considerations as follows. The $j=0$ moment
is the normalization integral, and is indeed equal to 1 by the condition (\ref{fi01})
\begin{eqnarray}
M_0 = \frac{1}{\hat P_{0,i+1}} f^{(i)}(1) = 1\,.
\end{eqnarray}
The first moment can be found again in closed form, and is equal to the forward
Libor rate $L_i^{\rm fwd}$, as expected 
\begin{eqnarray}
M_1 = \frac{1}{\hat P_{0,i+1}} \tilde L_i f^{(i)}(e^{\psi^2 t_i}) = 
\frac{\hat P_{0,i} - \hat P_{0,i+1}}{\hat P_{0,i+1}\tau_i} = L_i^{\rm fwd}\,.
\end{eqnarray}
This expresses the martingale condition for $L_i$ in the $\mathbb{P}_{i+1}$ measure.

More interesting is the result for the second moment $M_2$. This determines 
the equivalent lognormal volatility of the Libor rate $L_i$ as
\begin{eqnarray}\label{lnvol}
\sigma_{\rm LN}^2 t_i = \log\Big(\frac{M_2}{M_1^2}\Big) = \log
\Big(\hat P_{0,i+1} e^{\psi^2 t_i} \frac{f^{(i)}(e^{2\psi^2 t_i})}
{[f^{(i)}(e^{\psi^2 t_i})]^2}\Big)
\end{eqnarray}
In the small volatility limit $\psi^2 t_i \ll 1$ the ratio of generating functions
can be computed using the approximative formula (\ref{fi0}).
This gives 
\begin{eqnarray}\label{lnvolsmallpsi}
\sigma_{LN}^2 = \psi^2 ( 1 + O(\psi^2 t_i))\,.
\end{eqnarray}
This means that in the small volatility limit the caplet log-normal volatilities 
are approximately equal to $\psi$. This is useful for the calibration of the 
model, as the $\psi_i$ volatilities can be read off directly from ATM 
caplet volatilities.

When considered as a function of the volatility $\psi$, the moments 
$M_j, j\geq 2$ of the Libor probability distribution function $\Phi_i(L)$ have 
non-analytic dependence on $\psi$ at a value of the volatility 
$\psi_{\rm cr}^{(j)}$ given by the solution to the equation
\begin{eqnarray}\label{xst}
x_*(\psi) = e^{j\psi^2 t_i}
\end{eqnarray}
where $x_*(\psi)$ is the non-analyticity point of the generating function $f^{(i)}(x)$
at time horizon $t_i$. This is the point on the real positive axis where the
complex zeros of the generating function $f^{(i)}(x)$ pinch the real axis.
In general the position of the non-analyticity point $x_*(\psi)$
depends on the volatility parameter $\psi$, although it approaches  
a well-defined value in the very large volatility limit $\psi\to \infty$,
when  the generating function approaches the asymptotic expression
$f_\infty^{(i)}(x)$ given in Eq.~(\ref{fiasym}). The zeros 
and the non-analyticity point of the polynomial $f_\infty^{(i)}(x)$ have been 
studied in detail in \cite{PhaseTransition} for the case of a constant forward 
short rate.
As discussed above, approximating the generating function with its asymptotic
expression $f_\infty^{(i)}(x)$ leads to the result (\ref{r0cr}) for the 
critical volatility.

Assuming that $x_*(\psi)$ is independent on $\psi$ (as is the case for 
asymptotically large volatility),
from (\ref{xst}) it follows that the critical volatilities 
of the moments of the Libor distribution function are related as
\begin{eqnarray}\label{psisq}
\psi_{\rm cr}^{(j)} = \frac{\psi_{\rm cr}}{\sqrt{j}}\,,\quad \psi \to \infty
\end{eqnarray}
However, in reality the non-analyticity point occurs at moderate
values of the volatility $\psi$, for which $x_*$ has a pronounced 
dependence on $\psi$. This implies
that the simple relation (\ref{psisq}) is badly violated in practice.

We illustrate the non-analyticity in volatility of the moments $M_j$ on the
example of the second moment $M_2$. 
This is the most important moment from a practical point of view, as it 
determines the Black log-normal caplet volatility according to
Eq.~(\ref{lnvol}). In Figure~\ref{fig:ATMcapletvol} we show a plot of the 
equivalent 
Black caplet volatility $\sigma_{\rm LN}$ as a function of $\psi$ (red curve)
at the time horizon $i=30$ in a simulation with $n=40$ quarterly time steps. 

The equivalent log-normal volatility $\sigma_{\rm LN}$ has two turning
points, at $\psi$ around 0.3 and at 0.33. The critical point at the time
horizon considered here is $\psi_{\rm cr} = 0.33$, which corresponds to the
second point. In order to understand the first turning point, we show in
the Appendix the zeros of the generating function $f^{(i)}(x)$ together with 
two circles of radius $e^{\psi^2 t_i}$ and $e^{2\psi^2 t_i}$. From these 
plots one can see that the first turning point coincides with the zeros
crossing the larger circle of radius $e^{2\psi^2 t_i}$, and the second
turning point corresponds to the volatility $\psi$ at which the zeros cross 
the smaller circle, of radius $e^{\psi^2 t_i}$. 
Since the position of the zeros changes with $\psi$,
the first turning point (the critical volatility of the second moment $M_2$)
$\psi_{\rm cr}^{(2)} = 0.3$ differs from the large volatility limit
prediction following from Eq.~(\ref{psisq}) $\psi_{\rm cr}/\sqrt2 \simeq 0.23$, 
obtained by assuming stationary zeros. This illustrates the comment made above 
about the limited validity of Eq.~(\ref{psisq}).

Only the first few moments have non-analyticity points. The reason for this
is that at very low volatilities $\psi$, the zeros of the generating
function do not surround completely the origin, but a gap remains between
the real axis and the zeros. As the volatility increases, the zeros move
closer to the origin, and close together onto the real axis. However, 
at this point they have crossed already the circles of radii $e^{j\psi^2 t_i}$,
with $j > j_0$ such that the moments $M_j$ do not have a non-analyticity point 
for sufficiently large $j>j_0$. In other words, the equation (\ref{xst}) does not
have a solution for sufficiently large index $j$.
The maximal index $j_0$ of the moment of the
Libor pdf $L_i$ which still has a phase transition depends on the time 
horizon $t_i$ considered.

The price of an instrument which is
sensitive to the $j-$th moment will have a non-analyticity point at the 
corresponding value of the volatility. For the second moment this is the
case for example with the Libor payment in arrears, discussed in the next
section.

\subsection{Caplet pricing and Black caplet volatility}

A closed form expression for the 
caplet price can be found by direct evaluation of the expectation value in 
(\ref{Capletn})
\begin{eqnarray}\label{capletexact}
C_i(K) = P_{0,n} \sum_{j=0}^{n-i-1} c_j^{(i)}
[\tilde L_i e^{j\psi^2 t_i} N(f_1) - K N(f_2)]
\end{eqnarray}
with
\begin{eqnarray}
f_1 &=& - \frac{1}{\sqrt{t_i}} [x_0(K) - (j+1)\psi t_i]\\
f_2 &=& - \frac{1}{\sqrt{t_i}} [x_0(K) - j \psi t_i]\\
x_0(K) &=& \frac{1}{\psi}\log \frac{K}{\tilde L_i} + \frac12 \psi t_i\,.
\end{eqnarray}
This has the typical form of a mixing solution \cite{mixing,mixing2} for an
option price on an asset with a probability distribution consisting of a 
superposition of log-normal distributions. 

\begin{figure}
\begin{center}
\includegraphics[height=50mm]{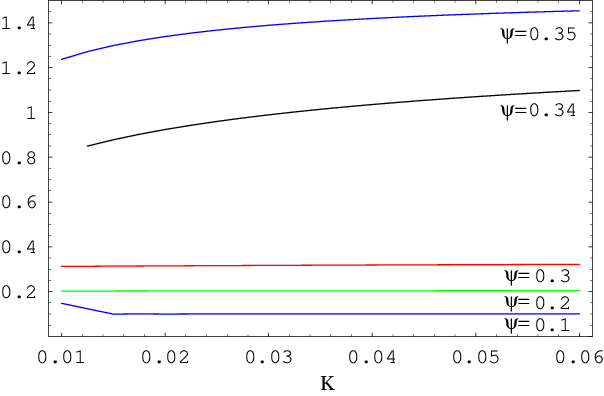}
\end{center}
\caption{Implied caplet volatility smile $\sigma_{\rm BS}(K)$ 
for several values of the volatility
$\psi$, as function of the strike $K$. The caplet is defined on the rate 
$L_{30}$ in a simulation with $n=40$ time steps, $\tau = 0.25$ and constant 
forward short rate $r_0=5\%$. The forward Libor is $L_i^{\rm fwd}=5.0\%$.}
\label{fig:capletskew}
\end{figure}

\begin{figure}
\begin{center}
\includegraphics[height=50mm]{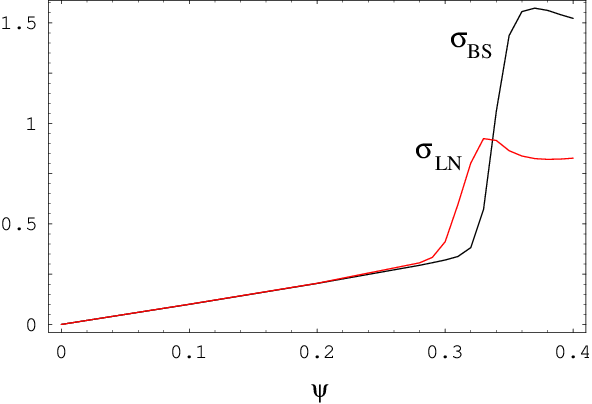}
\end{center}
\caption{The exact ATM caplet volatility $\sigma_{\rm BS}$ (black), and 
the equivalent log-normal caplet 
volatility $\sigma_{\rm LN}$ computed using (\ref{lnvol}) (red), as functions of the volatility 
parameter $\psi$.
The caplet strike is $K=5\%$, the forward Libor is $L^{\rm fwd}=5.0\%$,
corresponding to a constant forward short rate $r_0=5\%$. 
The remaining model parameters are $i=30$, $n=40$, $\tau = 0.25$.}
\label{fig:ATMcapletvol}
\end{figure}

Figure \ref{fig:capletskew} shows typical results for the Black (log-normal) 
caplet volatility $\sigma_{\rm BS}(K)$ for several values of the volatility $\psi$, 
obtained from the exact 
formula (\ref{capletexact}). At low values of $\psi$ the Black volatility is
independent of strike, which means that the distribution $\Phi_i(L)$
is aproximatively log-normal. In Figure~\ref{fig:ATMcapletvol} we show a 
plot of the exact ATM Black caplet volatility for ATM strike $K=5\%$.
From this plot one 
can see that, for small $\psi$, the ATM caplet volatility is to a very good 
approximation equal to $\psi$. This agrees with the prediction (\ref{lnvol}),
and confirms that for sufficiently small volatility, the Libor volatility
parameter $\psi$ is equal to a very good approximation with the caplet
volatility.
At larger values of $\psi$ above the critical volatility $\psi_{\rm cr}=0.33$,
the smile is not flat, which signals deviations from a log-normal distribution
for $\Phi_i(L)$.

In Figure~\ref{fig:ATMcapletvol} we show also the ATM
equivalent Black caplet volatility $\sigma_{\rm LN}$ given by 
Eq.~(\ref{lnvol}) (red curve), 
comparing it with the exact ATM caplet volatility $\sigma_{\rm BS}$ (black curve).
The critical volatility corresponding to the caplet shown in this plot is
$\psi_{\rm cr} = 0.33$. We observe that the exact and approximative
volatilities agree with each other for small $\psi$, where they satisfy
very well the approximative equality relation (\ref{lnvolsmallpsi}). This region
is the intended region of applicability of the model.

As the volatility $\psi$ is increased, a
sharp turn in the equivalent volatility $\sigma_{\rm LN}$ occurs at 
$\psi_{\rm cr}^{(2)} \sim 0.3$, which 
corresponds to the critical volatility of the second moment of the Libor
distribution function, as explained above.
The second turn point is at $\psi_{\rm cr} = 0.33$ which is the critical
volatility of the model at the maturity $t_i$ considered. It is interesting 
that for $\psi > \psi_{\rm cr}$ the
equivalent log-normal volatility decreases as the model volatility $\psi$
increases. 

The exact ATM caplet volatility has a first turning point which is closer 
to  the critical volatility $\psi_{\rm cr} = 0.33$. It starts to diverge 
from the equivalent log-normal volatility $\sigma_{\rm LN}$ at a lower
volatility $\psi \sim 0.3$, which thus is the point where the shape of the
Libor distribution function starts to deviate appreciably from a log-normal
shape. The fast increase
in the ATM caplet volatility above the critical volatility is explained by the
appearance of the long tail of the Libor distribution function $\Phi_i(L)$
extending to very large values of $L$. This gives a large contribution to the
caplet price, which is given by a simple integral over $\Phi_i(L)$
\begin{eqnarray}
C_i(K) = P_{0,i+1} \int_0^\infty dL (L-K)_+ \Phi_i(L) \,.
\end{eqnarray}
We remarked above on the numerical importance of the tail of the $\Phi_i(L)$ distribution
in relation to the integral (\ref{LiNA}), where it is needed in order for this
integral to reproduce its non-arbitrage value. Based on the same argument,
one expects that the
tail of this distribution will contribute significantly also to the 
caplet price $C_i(K)$ above the critical point.

\section{Libor payment in arrears}
\label{sec:4}

We consider in this section the pricing of a Libor payment in arrears in the
model with log-normally distributed rates in the terminal measure. 
We derive the convexity adjustment, and compare it
with the known convexity adjustment in the model with log-normal
caplet volatility. It will be seen that the convexity adjustment
in the model with log-normally distributed rates in the terminal
measure has a phase transition at two values of the Libor volatility,
in contrast to the latter, which is perfectly well-behaved as function
of the caplet volatility. 

The Libor payment in arrears pays the amount $L_i \tau_i$ at time $t_i$, 
where $L_i$ is the Libor rate for the $(t_i, t_{i+1})$ period, set at $t_i$. 
The price of this instrument in the terminal measure is
\begin{eqnarray}\label{Ai}
A_i &=& P_{0,n} \E_n[L_i(x_i)\tau_i P_{i,n}^{-1}(x_i)]
= P_{0,n} \E_n[L_i(x_i)\tau_i \hat P_{i,i+1}(x_i) (1 + L_i(x_i)\tau_i )]\,.
\end{eqnarray}
The first term, linear in $L_i$, is known exactly from the pricing of a forward
rate agreement 
\begin{eqnarray}
 \E_n[L_i(x_i)\tau_i \hat P_{i,i+1}(x_i) ] &=& \hat P_{0,i} - \hat P_{0,i+1}
=  P_{0,i+1} (L_i^{\rm fwd}\tau_i)\,.
\end{eqnarray}

The only non-trivial part is the pricing of the term quadratic in $L_i$, which can be expressed in terms of the
generating function
\begin{eqnarray}
\E_n[L^2_i(x_i)\hat P_{i,i+1}(x_i) ] = \tilde L_i^2 e^{ \psi^2 t_i} f^{(i)}(e^{2\psi^2 t_i})
\end{eqnarray}
Recall that the convexity-adjusted Libor is given by
\begin{eqnarray}
\tilde L_i = \frac{\hat P_{0,i} - \hat P_{0,i+1}}{f^{(i)}(e^{\psi^2 t_i}) \tau_i} = \hat P_{0,i+1} L_i^{\rm fwd} \frac{1}{f^{(i)} (e^{\psi^2 t_i})}
\end{eqnarray}
Combining all the pieces together we get for the price of a Libor payment in arrears
\begin{eqnarray}
\frac{A_i}{P_{0,i+1}} &=& (L_i^{\rm fwd}\tau_i) + 
\hat P_{0,i+1} (L_i^{\rm fwd}\tau_i )^2 e^{ \psi^2 t_i} 
\frac{ f^{(i)}(e^{2\psi^2 t_i})}{[f^{(i)} (e^{\psi^2 t_i})]^2} \,.
\end{eqnarray}
The second term is the convexity adjustment, and can be expressed in terms of the 
equivalent log-normal volatility $\sigma_{\rm LN}$ introduced above in (\ref{lnvol}).
We obtain the following result for the price of a Libor payment in arrears in the model 
with log-normally distributed Libors in the terminal measure
\begin{eqnarray}\label{APV}
A_i = P_{0,i+1} (L_i^{\rm fwd}\tau_i) \Big\{
1 +  (L_i^{\rm fwd}\tau_i ) e^{\sigma_{LN}^2 t_i} \Big\} \,.
\end{eqnarray}

This can be compared with the exact result for the price of a Libor payment in arrears
in a model with exact log-normal caplet volatility $\psi$ for the Libor $L_i$
\begin{eqnarray}\label{APVln}
A_i = P_{0,i+1} (L_i^{\rm fwd}\tau_i) \Big\{
1 +  (L_i^{\rm fwd}\tau_i ) e^{\psi^2 t_i} \Big\}
\end{eqnarray}
This model has a log-normal Libor distribution function in the
measure $\mathbb{P}_{i+1}$.
The result (\ref{APVln}) is identical with the price in the model with 
log-normal Libor in the terminal measure $\mathbb{P}_n$ (\ref{APV}), up to the
replacement $\sigma_{\rm LN} \to \psi$. 

At low volatility $\psi$, the equivalent volatility $\sigma_{\rm LN}$ 
is approximatively equal to $\psi$, see Eq.~(\ref{lnvolsmallpsi}).
For larger volatility $\psi$ it has a more complex behavior as discussed 
in Sec.~\ref{sec:3.1}, including two non-analyticity points at $\psi_{\rm cr}^{(2)}$ and 
$\psi_{\rm cr}$, as observed in Fig.~\ref{fig:ATMcapletvol}.
This means that the price of this instrument has the same non-analytical 
behavior in $\psi$ as $\sigma_{\rm LN}(\psi)$.

Similar non-analyticity effects can be expected to appear in the
pricing of other interest rates derivatives, and are introduced either through
non-analytic behavior in the convexity-adjusted Libors $\tilde L_i$, or 
through the moments of the Libor distribution function in the forward measure
$\mathbb{P}_{i+1}$. Thus non-analyticity effects appear to be a 
generic feature of models with log-normally distributed rates in the
terminal measure.

\section{Conclusions}
\label{sec:5}

We considered in this paper the dynamics of an 
interest rate model with log-normally distributed rates in the terminal 
measure. Such models are used in financial practice as particular parametric
realizations of the Markov functional model, and as approximations to models
with log-normal caplet smile, such as the log-normal Libor market model. 
Using the exact solution of the model we studied the dependence on volatility
of the distributional properties of the dynamical quantities of the model 
and their implications for pricing interest rate derivatives. The main result
of the study is the existence of a previously unobserved sharp transition
at a critical value of the volatility.
Above the critical volatility certain expectation values and convexity
adjustments have an explosive growth.
The values of the critical volatility in simulations with 10-30y and interest 
rates around 5\% are comparable with actual log-normal caplet volatilities 
observed in the market, such that the existence of this phase transition is of 
practical relevance, and imposes a limit on the applicability of the model.

It has been long known that models with log-normally distributed rates suffer
from singular behavior. This was observed in \cite{HW,SS} 
in the context of the Dothan model, and of the Black-Karasinski model. 
However, the
phenomenon discussed here appears to be different in several respects: first,
the singularity discussed in \cite{HW,SS} was shown to appear for a continuous time
model, while the model considered here is defined in discrete time. Second, the
transition discussed here appears at a well-defined finite value of the volatility,
while the divergence studied in \cite{HW,SS} is independent of volatility.

The results of this paper show that at low volatilities
a log-normal caplet smile can be well reproduced by assuming Libor 
log-normality in the terminal measure; however at larger volatility this 
property is not preserved, and a non-trivial cap smile is generated.
These results spell out the
limits of applicability of the log-normal parameterization (\ref{Lipdf}) 
for describing
an interest rate market with log-normal caplet smile. Such models can
be applied only for sufficiently low caplet volatility, below the 
critical volatility.

The underlying reason for this limitation is a change in the shape of
the probability distribution function of the Libor rates in their forward 
measure around the critical volatility. For small volatility  the Libor
pdf has a typical humped shape, centered around the forward Libor value.
However, at the critical volatility this pdf changes suddenly, and
it collapses to small Libor values, in addition to developing a long tail. 
Furthermore, the moments of this probability distribution function have also 
sharp transitions as functions of volatility. 

\begin{figure}[b]
\centering
\begin{tabular}{cc}
\includegraphics[width=2.5in]{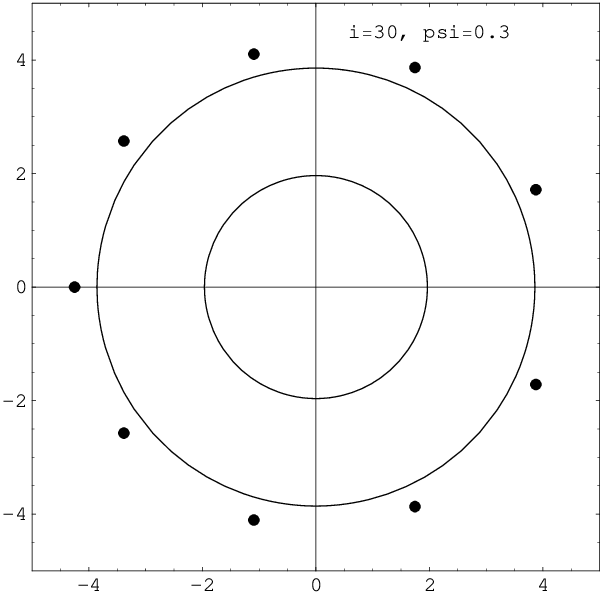} &
\includegraphics[width=2.5in]{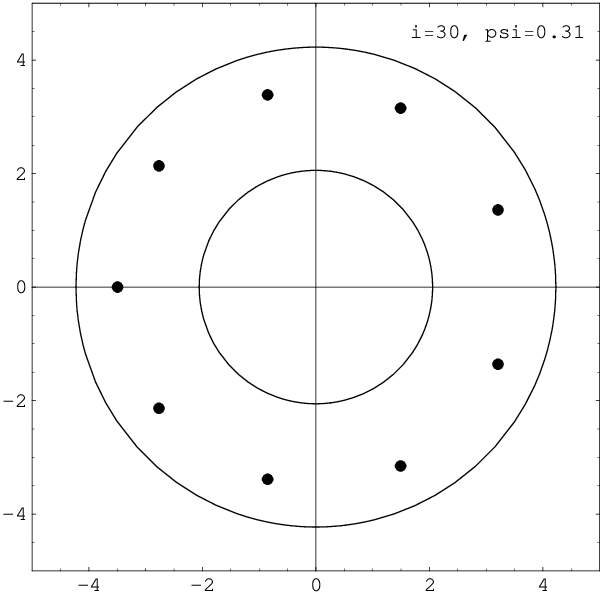} \\
\includegraphics[width=2.5in]{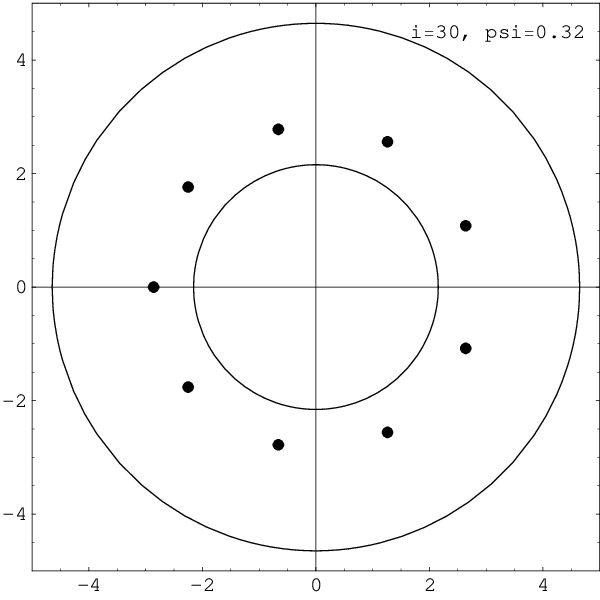} &
\includegraphics[width=2.5in]{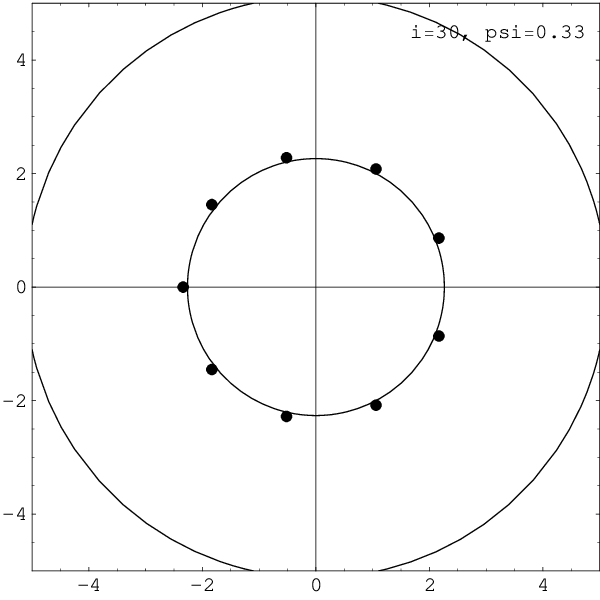} 
\end{tabular} 
\caption{The position of the zeros of the generating function $f^{(i)}(x)$
for several values of the volatility $\psi = 0.3-0.33$ at the time horizon
$i=30$ in a simulation with $n=40$ quarterly time steps ($\tau=0.25$). The
two circles shown have radii $e^{\psi^2 t_i}$ and $e^{2\psi^2 t_i}$. The
zeros cross these circles at $\psi_{\rm cr}$ and $\psi^{(2)}_{\rm cr}$, 
corresponding to the critical volatility of the model, and to the critical
volatility of the second moment of the Libor distribution function, respectively.}
\label{fig:zeros}
\end{figure}

These phenomena have implications for interest rate derivative pricing under the
given model, and we considered as 
concrete examples caplets on Libor rates and Libor payments in arrears. 
The caplet prices, the Black caplet volatilities, and the convexity
adjustment for Libor payments in arrears display also
sharp transitions as functions of the model volatility. Based on these examples,
it is plausible to expect that this phenomenon occurs also for other interest rate 
derivatives, and is a general feature of the interest rates models with log-normally 
distributed rates in the terminal measure. 

The effects discussed in this paper are due to a contributions to the
expectation values from a region in the state variable (Markovian driver) which
is usually assumed to be unimportant in practice, as it is associated with
very large interest rates $\sim 100\%$. This region is usually truncated off
in tree and finite difference simulations, or is very poorly sampled in
Monte Carlo simulations, unless extremely  high numbers of paths are used. 
This implies that usual numerical implementations of this model 
do not capture correctly the behavior of the model in the large volatility
phase, and thus the phase transition is not visible under these simulation 
methods. In practice one can take the view that the implementation version of the 
model with its built-in limitations (e.g. limits on the range of the Markovian 
driver $x(t)$) {\em is} the model. This corresponds to a truncation of the 
original model, and the numerical consistency of this truncation must be carefully
verified. 

The arguments of this paper are limited to the time-homogeneous setting 
of a uniform volatility, but it is plausible that a similar phenomenon will occur 
also in the practically relevant but analytically more complex case of time-dependent 
volatility $\psi_i$. 
This is confirmed by the study of a related model in \cite{LatticeGas},
where the stochastic driver $x(t)$ is replaced with an Ornstein-Uhlenbeck process.
This corresponds to the Black-Karasinski model in  the terminal measure,
with mean reversion and term structure of caplet volatilities. The exact solution 
of this model obtained in \cite{LatticeGas} shows the presence of a similar
phase transition in volatility in the convexity adjustments of the model. 
Finally, it would be interesting to investigate whether some of these results
persist also in a model with exact log-normal caplet volatility, such as
the exact Markov functional model \cite{BH,HKP} or the Libor market model \cite{BGM}. 
This is plausible in view of the result obtained in \cite{Gerhold}, according to which 
the Libor distribution function in the terminal measure in the LMM has log-normal tails,
which is similar to the distributional property (\ref{Lipdf}) of the model considered here.

\bigskip

{\bf Acknowledgements.} I am grateful to Emanuel Derman and the participants 
at the Columbia University IEOR seminar for comments and discussions, and to an
anonymous referee for useful comments and criticism. The information, views and
opinions set forth in this publication are those of the author, and are in no
ways sponsored, endorsed, or related to the business of J.~P.~Morgan Chase
\& Co. (``J.~P.~Morgan''). J.~P.~Morgan does not warrant the publication's
completeness or accuracy, and makes no representations regarding the use of the
information set forth in this publication.
In no event shall J.~P.~Morgan be liable for any direct, indirect,
special, punitive, or consequential damages, including loss of principal and/or
lost profits, even if notified of the possibility of such damages. Nothing in this
publication is intended to be an advertisement or offer for any J.~P.~Morgan
service.

\section*{Appendix}

We illustrate in this Appendix the relation between the position of the
zeros of the generating function $f^{(i)}(x)$ and the non-analyticity properties of the
moments of the Libor distribution function discussed in Sec.~\ref{sec:3.1}. 
We take as a concrete example a simulation with $n=40$ quarterly time
steps, with constant forward short rate $r_0=5\%$, and we examine the zeros of the
$f^{(i)}(x)$ at the time step $i=30$. 

The plots in Fig.~\ref{fig:zeros} show the movement of the zeros of
$f^{(30)}(x)$ as a function of the volatility $\psi$ for $\psi=0.3-0.33$.
On the same plots are shown also two circles with radii  
 $e^{\psi^2 t_i}$ and $e^{2\psi^2 t_i}$. The values of the volatility 
at which the zeros cross these circles are the critical volatility
$\psi_{\rm cr}$ and the critical volatility of the second moment
$\psi_{\rm cr}^{(2)}$, respectively. These critical volatilities are 
visible as turning points in the plot of the equivalent log-normal
volatility $\sigma_{\rm LN}$ as function of $\psi$ in Fig.~\ref{fig:ATMcapletvol}.

A similar picture holds for the higher order moments. For example, the $j$-th
moment of the Libor distribution function $M_j = \E_{i+1}[L_i^j]$ will have
a non-analyticity point at $\psi_{\rm cr}^{(i)}$. This corresponds to that
value of the volatility where the zeros of $f^{(i)}(x)$ cross the 
circle of radius $e^{j\psi^2 t_i}$. As mentioned in the text, for
sufficiently high order moments the zeros do not surround completely the
origin, and these moments will not have a phase transition.

\end{document}